\documentclass[12pt]{article}
\usepackage[dvips]{graphicx}
\usepackage{subfigure}

\usepackage{epsfig,amsmath,amssymb,verbatim,mathrsfs}

\def\beq{\begin{eqnarray}}
\def\eeq{\end{eqnarray}}
\def\bea{\begin{eqnarray}}
\def\eea{\end{eqnarray}}

\newcommand{\ir}{\text{\tiny IR}}
\newcommand{\uv}{\text{\tiny UV}}
\newcommand{\cft}{\text{\tiny CFT}}

\newcommand{\be}{\begin{equation}}
\newcommand{\ee}{\end{equation}}

\begin{document}

\setlength{\baselineskip}{0.2in}


\begin{titlepage}
\noindent
\flushright{March 2012}
\vspace{0.2cm}

\begin{center}
  \begin{Large}
    \begin{bf}
Sommerfeld Enhancement from Multiple Mediators
     \end{bf}
  \end{Large}
\end{center}

\vspace{0.2cm}

\begin{center}

\begin{large}
{Kristian L.~McDonald}\\
     \end{large}
\vspace{0.5cm}
  \begin{it}
Max-Planck-Institut f\"ur Kernphysik,\\
Saupfercheckweg 1, 69117 Heidelberg, Germany\\\vspace{0.5cm}
\vspace{0.3cm}
kristian.mcdonald@mpi-hd.mpg.de
\end{it}
\vspace{0.5cm}

\end{center}


\begin{abstract}
We study the Sommerfeld enhancement experienced by a scattering object that couples to a tower of mediators. This can occur in, e.g., models of secluded dark matter when the
mediator scale is
generated naturally by hidden-sector confinement. Specializing to the case of a
confining CFT, we show that off-resonant values of the enhancement can be increased by $\sim20\%$ for cases of interest when (i) the (strongly-coupled) CFT admits a weakly-coupled dual description and (ii) the conformal symmetry holds up to the Planck scale. Larger enhancements are possible for lower UV scales due to an increase in the coupling strength of the tower.

\end{abstract}

\vspace{1cm}

\end{titlepage}

\setcounter{page}{1}


\vfill\eject


\section{Introduction}
In recent years there has been much interest in the Sommerfeld
enhancement that arises when a scattering object is coupled to a light
mediator. This enhancement can increase the cross section for
scattering/annihilation processes in a velocity-dependent fashion (see, e.g.,~\cite{Cassel:2009wt,Slatyer:2009vg,Chen:2009ch,Feng:2010zp} and references therein). In this work we consider the Sommerfeld enhancement experienced by a scattering object that is coupled to a tower of mediators. This can occur in, e.g., models of secluded dark matter~\cite{Pospelov:2007mp,Finkbeiner:2007kk,Fayet:2007ua} when a hierarchically small mediator-scale is generated naturally by hidden-sector confinement~\cite{bh_km}. As a specific example, we consider a scattering object that is charged under a weakly-gauged global $U(1)$ symmetry of a (broken) CFT. We show that the enhancement can increase as a result of multi-mediator exchange.

The basic point is straight forward. If the scattering field (here called $\psi$, with mass $M$) is coupled to multiple mediators ($\phi_n$, with masses $m_n\ll M$), the exchange of this ``tower'' of mediators can modify the Sommerfeld enhancement relative to the standard single-mediator result. It is easy to see how such a situation could arise. Consider the case where $\psi$ is charged under a $U(1)$ symmetry, with the corresponding gauge boson ($\gamma'$) playing the role of Sommerfeld-mediator. Now let the $U(1)$ factor also weakly-gauge a global symmetry of a strongly-interacting sector that confines in the infrared (IR) at a scale $\Lambda_{\ir}\ll M$. The confinement is such that $U(1)$ is broken during the phase transition and the mass of the corresponding gauge boson is set by the confinement scale, $m_{\gamma'}\sim \Lambda_{\ir}$.  One can think of the confining sector as providing a technically natural (i.e. radiatively stable) way to generate the hierarchy $m_{\gamma'}\ll M$. 
If the strongly-interacting sector contains spin-one states, these may mix with $\gamma'$, much like the photon mixes with the $\rho$-meson in the Standard Model (SM). The lightest spin-one states will typically have masses on the order of $\Lambda_{\ir}$, and thus the mixing induces effective interactions between $\psi$ and multiple light mediators. These extra mediators can modify the Sommerfeld enhancement.

The study of such a system is complicated by the invocation of a strongly-interacting sector. Fortunately, however, via the AdS/CFT correspondence~\cite{Maldacena:1997re,ArkaniHamed:2000ds}, one can construct weakly-coupled dual theories for classes of strongly-interacting 4D models. Using this approach one can construct models in which a fundamental (or non-composite) field $\psi$ is charged under a weakly-gauged $U(1)$ symmetry of a strongly-interacting conformal sector. When the  conformal symmetry is broken  at a scale $\Lambda_{\ir}\ll M$, such that the $U(1)$ symmetry is also broken, one arrives at a calculable model with multiple mediators. More specifically, the Lagrangian for the dual 4D theory of a bulk vector in a slice of $AdS_5$ contains the terms
\bea
\mathcal{L} \  =\  \mathcal{L}_{\psi}\ +\ \mathcal{L}_{\cft}\ +\  \mathcal{L}_{\gamma'}\ +\  A_\mu'\mathcal{O}^\mu_{\cft} \ +\  \dots,
\eea
where $\mathcal{O}^\mu_{\cft}$ creates spin-one states in the conformal sector. Observe that the last term mixes $\gamma'$ with the spin-one states of the CFT --- if $\psi$ has non-zero $U(1)$ charge it thus acquires a coupling to the CFT modes.\footnote{At energies $E\gg \Lambda_{\ir}$ the operator $\mathcal{O}^\mu_{\cft}$ describes the CFT in the conformal regime, while at energies $\lesssim \Lambda_{\ir}$ it describes the discrete composite spin-one states.} 

Our main interest is in studying the Sommerfeld enhancement due to a tower of mediators. To understand our analysis one need not be familiar with the details of warped 5D models, nor the AdS/CFT dictionary. Instead it suffices to consider an effective action of the form
\bea
S\supset\int d^4x \left\{i\bar{\psi}\gamma^\mu \partial_\mu \psi
  +\sum_n e_{n}\bar{\psi}\gamma^\mu \psi \phi_\mu^{n}- M\bar{\psi}\psi\right\},\label{DM_action_tower_intro}
\eea  
in which the tower of mediators, labeled by $n\ge0$, have ordered masses ($m_n>m_{n'}$ for $n>n'$) and couplings ($\alpha_n\equiv e_n^2/4\pi<\alpha_{n'}$ for $n>n'$).  As we will show, this scenario can be motivated by considering weakly-coupled dual 5D theories, which generate specific relations among the masses $m_n$, and the couplings $\alpha_n$. 

Although our purpose is to study this system in a general sense, we
note that much of the recent interest in the Sommerfeld enhancement
has related to experiments like
PAMELA~\cite{Adriani:2008zr},
\emph{Fermi}~\cite{Abdo:2009zk,Ackermann:2010ij} and
ATIC~\cite{:2008zzr,Panov:2011zw}. These experiments observe an unexpected
increase in the cosmic-lepton signal at energies $\gtrsim10$~GeV,
suggesting a new source of high energy electrons and positrons. It was conjectured that this lepton excess could be an indirect signal of WIMP Dark Matter (DM) annihilations, as occurs in models where DM annihilates into a light mediator~\cite{ArkaniHamed:2008qn}. Typically, the cross section required to explain the lepton signal is larger than the expected freeze-out cross section by a factor of $\sim\mathcal{O}(10^2)$. However, the mediator increases the present-day low-velocity annihilation cross section via the Sommerfeld effect, enabling models where the DM achieves thermalization in the early universe and yet annihilates to (potentially) produce the present-day lepton signal. Though it is not our intention (or focus) to construct models along this line, we briefly comment on the use of our results within this context. 

A more detailed analysis of DM coupled to a tower of mediators will appear elsewhere~\cite{bh_km}. The results of the present paper, for the Sommerfeld enhancement from a tower of mediators admitting a dual description, are necessary in order to understand the cosmic ray signal in the companion paper. Also note that the enhancement due to multiple mediators could also be of interest in more general frameworks. For example, when a scattering object is charged under a $U(1)$ factor that kinetically mixes with multiple other $U(1)$ factors~\cite{Heeck:2011md}, or when a $U(1)$ symmetry is broken by a strongly-coupled sector that does not admit a dual description. In such cases the coupling and mass relations among mediators will differ, though our basic methodology will apply.

The layout of this paper is as follows. In Section~\ref{sec:single_med} we briefly review aspects of the standard (single-mediator) Sommerfeld enhancement. Sections~\ref{sec:Somm_tower_no_spilt} and~\ref{sec:split_sommer_tower} contain our discussion of the Sommerfeld enhancement for a tower of mediators, in the elastic (Dirac), and inelastic (Majorana), cases respectively. A weakly-coupled dual model is presented in Section~\ref{sec:dual_model}. We note that familiarity with Section~\ref{sec:dual_model} is not necessary in order to follow the earlier sections; the latter affect the former only via the particular relations employed for $m_n$ and $\alpha_n$. Some brief comments in relation to the cosmic-lepton excess appear in Section~\ref{sec:cosmic_lepton}.  The paper concludes in Section~\ref{sec:conc} and more details regarding the dual model appear in an Appendix.
\section{The Standard Enhancement from a Single Mediator\label{sec:single_med}}
In the sections that follow we obtain the Sommerfeld enhancement for a 4D fermion coupled to a tower of vector mediators. As a helpful point of reference, we briefly discuss aspects of the standard Sommerfeld enhancement for a scattering object of mass $M$ coupled to a mediator $\phi$. This serves to establish our notations, to remind the reader of the key features of the standard analysis, and to make the presentation more self-contained. Our discussion follows Refs.~\cite{ArkaniHamed:2008qn,Slatyer:2009vg}.

The Sommerfeld enhancement $S$ from a central potential $V(r)$ is found by solving the following  Schr\"odinger equation,
\bea
-\frac{1}{2M}\frac{d^2\chi}{dr^2}\ +\ V(r)\chi =\frac{p^2}{2M}\chi.\label{sommerfeld_schrodinger}
\eea
We first consider a Coulomb potential, $V(r)=-\alpha/2r$. Changing to the 
variable $x=\alpha M r$, and defining $\epsilon_v=v/\alpha$, gives
\bea 
\chi''+x^{-1}\chi=-\epsilon_v^2\chi,\label{coulombschrodinger}
\eea
This equation can be solved, subject to
the boundary conditions $\chi(0)=1$, and $\chi'(x)=i\epsilon_v \chi(x)$ for
$x\rightarrow\infty$. The Sommerfeld enhancement is then
given by $S=|\chi(\infty)|^2$. For the case of a Coulomb potential an analytic solution is possible:
\bea
S=\left|\frac{\pi/\epsilon_v}{1-\exp(-\pi/\epsilon_v)}\right|,
\eea
and the limiting values for the enhancement are:
\bea
S\rightarrow
\left\{
\begin{array}{cccc}
1& &\mathrm{for}&\epsilon_v\gg 1\\
\pi/\epsilon_v& &\mathrm{for}& \epsilon_v\ll 1
\end{array}.
\right. \label{coulomb_S_limits}
\eea
The latter result is helpful in what follows. 

\begin{figure}[ttt]
\begin{center}
        \includegraphics[width = 0.7\textwidth]{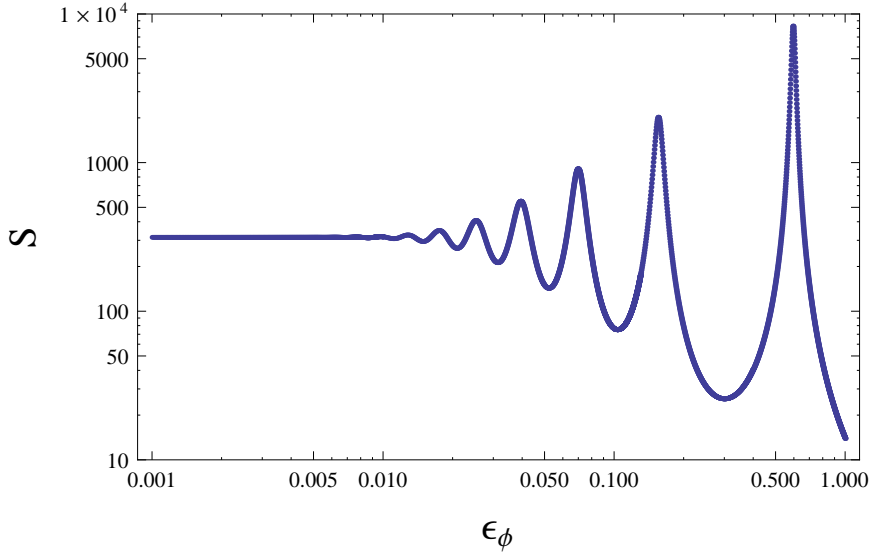}
\end{center}
\caption{The Sommerfeld enhancement for a Yukawa potential, as a function of the parameter
  $\epsilon_{\phi}=m/\alpha M$ for the fixed value 
  $\epsilon_v=0.01$.}\label{sommerfeld_standard_yukawa}
\end{figure}

Next,  consider a Yukawa potential:
\bea
V(r)=-\frac{\alpha}{2r}e^{-m r},
\eea
where $m$ is the mediator mass. Changing variables as before gives:
\bea
 \frac{d^2\chi}{dx^2}\ +\ \frac{e^{-\epsilon_\phi x}}{x}\chi
 =-\epsilon_v^2\chi\label{yukawa_sommerfeld_eq}\ ,
\eea
where the dimensionless parameters are
\bea
\epsilon_v=\frac{v}{\alpha}\ ,\quad\epsilon_\phi=\frac{m}{\alpha M}.
\eea
In these units the range of the potential is $(\epsilon_\phi)^{-1}$. Eq.~(\ref{yukawa_sommerfeld_eq}) can be solved numerically, subject to the boundary conditions $\chi(0)=1$ and $\chi'(x)=i\epsilon_v \chi(x)$ for
$x\rightarrow\infty$. We plot the Sommerfeld enhancement, $S=|\chi(\infty)|^2$, due to an attractive Yukawa
potential in
Figure~\ref{sommerfeld_standard_yukawa}. The fixed value
$\epsilon_v=0.01$ is used in the plot and the enhancement is shown as a function of $\epsilon_\phi$. There are two key features to observe. Firstly, the enhancement saturates for small values of $\epsilon_\phi\ll\epsilon_v$. In this case the Yukawa potential is well approximated by its leading term and behaves effectively like a Coulomb potential, giving $S\rightarrow \pi/\epsilon_v$, in accordance with \eqref{coulomb_S_limits}. Secondly, the enhancement displays resonances in the regime $\epsilon_v\ll\epsilon_\phi$. Semi-analytic results show that these resonances occur at $\epsilon_\phi\simeq 6/(\pi^2n^2)$ for integer $n>0$, giving $S\simeq (\pi^2/6) (\epsilon_\phi/\epsilon_v^2)$~\cite{Cassel:2009wt,Slatyer:2009vg}. Thus, the enhancement goes like $1/\epsilon_v^2$ on-resonance. Also note that the enhancement turns off for $\epsilon_\phi\gtrsim 1$. 
\section{The Enhancement from a Tower of Mediators\label{sec:Somm_tower_no_spilt}}

We are interested in the Sommerfeld enhancement for a scattering
object that couples to a tower of mediators. Label the tower by integer $n\ge0$, such that the masses and couplings are
ordered with $m_n>m_{n'}$, and $\alpha_n<\alpha_{n'}$, for
$n>n'$. The
latter condition need not hold for an arbitrary tower of mediators, but 
holds for a tower admitting a weakly-coupled dual
description (see Section~\ref{sec:dual_model}). In this case, the masses
are related via
\bea
m_0\lesssim \Lambda_\ir\qquad\mathrm{and}\qquad m_n\sim n\times\Lambda_\ir\quad\mathrm{for}\quad
n\ge1,
\eea
where $\Lambda_\ir$ is an IR  mass-gap/confinement-scale. Furthermore, the couplings are related as
$\alpha_n/\alpha_0\sim 1/n$ for $n>1$ (see \eqref{DM_coupling_n_}). We use this case as a concrete
example but otherwise aim to keep the discussion general. 

The Sommerfeld enhancement from the tower of mediators follows from employing a sum over Yukawa
potentials:
\bea
V(r)&=&\sum_nV_n=-\sum_{n=0}\ \frac{\alpha_n}{2r}\ e^{-m_n r}.
\eea
In the limit that the modes with $n>0$ decouple, the lightest mode ($n=0$) serves as a standard mediator. We thus change variables to dimensionless quantities defined in terms of the mass and coupling of the lightest mediator:
\bea
\epsilon_{\phi}=\frac{m_0}{\alpha_0
  M}\ ,\quad \epsilon_v=\frac{v}{\alpha_0}\quad \mathrm{and}\quad
x=\frac{r}{(\alpha_0M)^{-1}}\ .
\eea 
In terms of these, the Schr\"odinger equation becomes
\bea
 \frac{d^2\chi}{dx^2}\ +\ \frac{1}{x}\left\{e^{-\epsilon_{\phi}
     x}+\sum_{n>0}\frac{\alpha_n}{\alpha_0}e^{-\delta_n \epsilon_\phi x}\right\}\chi
 =-\epsilon_v^2\chi\label{KK_yukawa_sommerfeld}\ ,
\eea
 This equation can again be solved numerically to determine the
 Sommerfeld enhancement. 

With the couplings and masses ordered as
 above, one can, in general, cut the sum off at some value
 $n_{\max}$. The reason for this is readily understood. The standard
 enhancement turns off for $\epsilon_\phi\gtrsim1$. For a
 tower of mediators such that
 $\epsilon_\phi<1$, the zero-mode enhancement is
 ``turned on'' and there will, in general, exist a value of $n=n_{\max}$ such that
 $\delta_{n}\epsilon_\phi\gtrsim 1$ for
 $n>n_{\max}$. Thus, for $n>n_{\max}$ the enhancement from the $n$-th
 mode turns off. For the specific case of a tower admitting a
 weakly-coupled dual description, we have $m_0\sim \Lambda_{\ir}$ and
 $m_n\sim n\times \Lambda_{\ir}$
($n>0$), so that $\delta_n=m_n/m_0\sim n$. Thus, we can cut the sum
off at $n_{\max}\sim (\epsilon_\phi)^{-1}$. Physically, this is equivalent to the statement that the
non-relativistic scattering of particles with mass $M$ is insensitive
to short distance physics at scales  $r\ll M^{-1}$, and such heavy
modes can be integrated out. 

For $m_n>m_0$ and $\alpha_n<\alpha_0$,  the potentials $V_{n}$ for
$n>0$ are subdominant to the zero-mode potential 
($V_0$) at large $x$. The long-distance behavior of the solution is
therefore governed by the lightest mode and $V\simeq V_0$ for $x\gg
1$, as one would expect. On the other hand, when $\epsilon_\phi\ll1$
the exponentials can become unimportant. Indeed, for $\epsilon_\phi\ll \epsilon_v$ one can approximate the zero-mode potential by a Coulomb potential, $V_0\simeq \alpha_0/2r$. In this region of parameter space the contribution from the higher modes is
\bea
\frac{1}{x}\ \sum_{n>0} \ \frac{\alpha_n}{\alpha_0}\ e^{-\delta_n \epsilon_\phi x}&\simeq &\sum_{n=10}^{n_{\max}}\frac{\alpha_n}{\alpha_0}\left\{\frac{1}{x} +\delta_n\epsilon_\phi\right\}\ +\dots,
\eea
where the dots denote modes with
$n>n_{\max}$. Noting that
\bea
\frac{\alpha_n}{\alpha_0}\ \delta_n\epsilon_\phi& \sim& \frac{\epsilon_\phi}{\sqrt{\log (\Lambda_{\uv}/\Lambda_{\ir})}}\ \lesssim\ \epsilon_\phi,
\eea
for the dual model (see Section~\ref{sec:dual_model}),  where $\Lambda_{\uv}$ is the UV cutoff for which we take $\Lambda_{\uv}\sim M_{Pl}$, we learn the following. In the regime $\epsilon_\phi\ll\epsilon_v$, when the zero-mode potential is well approximated by a Coulomb potential, the potentials $V_n$ for higher modes with $\delta_n\epsilon_\phi<1$ can also be approximated by Coulomb potentials. Using this approximation in the Schr\"odinger equation one finds
\bea
 \frac{d^2\chi}{dx^2}\ +\ \frac{1}{x}\left\{1
    +\sum_{n=1}^{n_{\max}}\frac{\alpha_n}{\alpha_0}\right\}\chi
 \simeq-\epsilon_v^2\chi,\quad\mathrm{for}\quad \epsilon_\phi\ll\epsilon_v\label{KK_yukawa_sommerfeld_approx}\ .
\eea
 Upon defining an effective coupling,
\bea
\bar{\alpha} \equiv \sum_{n=0}^{n_{\max}}\alpha_n\ ,
\eea
and changing variables to  $\bar{x}= \bar{\alpha} x/\alpha_0$, we rewrite
\eqref{KK_yukawa_sommerfeld_approx}
as
\bea
 -\frac{d^2 \chi}{d\bar{x}^2}\ -\ \frac{1}{\bar{x}}\ \chi
 =\bar{\epsilon}_v^2\chi\label{KK_coulomb_sommerfeld_eq}\ ,
\eea
where $\bar{\epsilon}_v=v/\bar{\alpha}$. Comparing with \eqref{coulombschrodinger} and using \eqref{coulomb_S_limits} we immediately deduce the enhancement in the limit
$\epsilon_\phi\ll\epsilon_v\ll1 $,
\bea
S\rightarrow \frac{\pi}{\bar{\epsilon}_v}=\frac{\pi}{\epsilon_v}\times \left\{\sum_n^{n_{\max}}\frac{\alpha_n}{\alpha_0}\right\}.\label{small_ephi_limit}
\eea
We will show in Section~\ref{sec:dual_model} that one can approximate the sum, giving 
\bea
S&\simeq& \frac{\pi}{\epsilon_v}\times\left\{1+ \frac{\log(1/\epsilon_\phi)}{\log(\Lambda_{\uv}/\Lambda_{\ir})}\right\}\ ,\label{inelastic_saturated_enhancement}
\eea
for the small-$\epsilon_\phi$ enhancement. Observe that the deviation from the standard (single mediator) small-$\epsilon_\phi$ result increases with decreasing $\epsilon_\phi$. For example:
\bea
S&\rightarrow &\frac{\pi}{\epsilon_v}\times \left\{
\begin{array}{cccc}
1.11&&\mathrm{for}&\epsilon_\phi\sim10^{-2}\\
1.16&&\mathrm{for}&\epsilon_\phi\sim10^{-3}
\end{array}
\right. \ ,\label{sommer_planck_rough}
\eea
with $\epsilon_\phi\ll\epsilon_v\ll1$ and $\Lambda_{\uv}\sim M_{Pl}$. Thus, relative to the standard result the enhancement is increased by a factor of
$\sim15\%$ for $\Lambda_{\uv}\sim M_{Pl}$. A more careful numerical treatment, gives
$S\rightarrow (\pi/\epsilon_v)\times 1.2$ for $\epsilon_\phi=10^{-3}$,
in agreement with the estimate.\footnote{As we show in Section~\ref{sec:dual_model}, the estimate is smaller than
  the exact result as it neglects the $n$-dependence of the logarithms that appear in $\alpha_n$, and thus slightly underestimates the
  couplings $\alpha_n$ for $n>0$.} 

\begin{figure}[ttt]
\begin{center}
             \includegraphics[width = 0.7\textwidth]{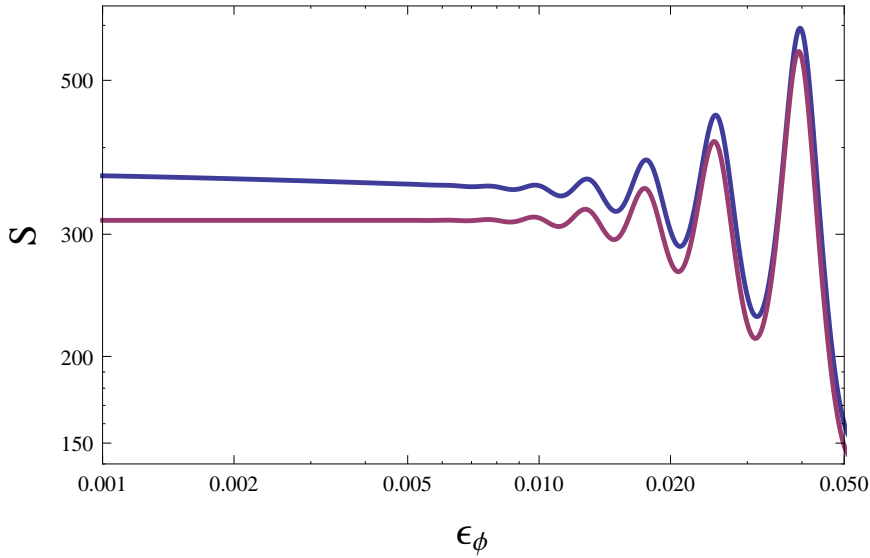}
\end{center}
\caption{Sommerfeld enhancement as a function of the parameter
  $\epsilon_{\phi}=m_0/\alpha_0M$ for the fixed value 
  $\epsilon_v=0.01$. The upper (blue) curve is for a tower of mediators; the lower (purple) curve
 is for a single mediator with
  coupling $\alpha_0$ and mass
   $m_0$. The two curves deviate in the small $\epsilon_\phi$ regime
   while for larger $\epsilon_\phi$ the effects of the heavier
   mediators decouple. }\label{zoom_sommerfeld_standard_yukawa}
\end{figure}

The Sommerfeld enhancement due to the tower, along with the pure
zero-mode result, is plotted as a function of $\epsilon_\phi$
in Figure~\ref{zoom_sommerfeld_standard_yukawa}
 for the fixed value
$\epsilon_v=0.01$ (with $\Lambda_{\uv} \sim M_{Pl}$). The latter result is equivalent to the standard enhancement from a single
mediator with mass $m_0$ and coupling $\alpha_0$. Observe that the discrepancy
between the standard result (lower curve), and the enhancement from the
tower (upper curve), increases with decreasing $\epsilon_\phi$. This is because more members of the tower contribute
to the enhancement as one decreases $\epsilon_\phi$. For fixed
$\alpha_0$, $\epsilon_\phi$ decreases with decreasing $\Lambda_\ir$
--- a limit that results in more members of the tower having mass $m_n<M$, and therefore
increases the number of modes that contribute to the enhancement.\footnote{Note that the present
  analysis cannot be trusted for arbitrarily small $\epsilon_\phi$. If
  $M$ gets too large ($M\gtrsim\Lambda_\uv$), the gravity description in the dual theory
  breaks down and the back reaction of $\psi$ should be
  included. Furthermore, the limit $\Lambda_\ir\rightarrow0$ removes
  the mass gap of the CFT, and the mediator $\gamma'$ becomes
  non-dynamical as its coupling runs to zero in the IR~\cite{ArkaniHamed:2000ds}. For $M\lesssim
  \Lambda_\uv$ and
  finite $\Lambda_{\ir}$, the present analysis, based on a dual description, can be trusted.}  Also note that the two curves agree for larger values of
$\epsilon_\phi$, demonstrating the insensitivity to the heavier
mediators in this limit. For values of
$\epsilon_\phi$ even larger than appears in the figure, the difference
between the standard single-mediator result and the tower is negligible, and in  both cases the enhancement turns off completely for $\epsilon_\phi \gtrsim 1$. Also note that the location of the resonances is not significantly affected by the tower, due to the dominant coupling of the lightest mode. In models where the couplings are not ordered as $\alpha_n<\alpha_{n'}$ for
$n>n'$, this feature would not, in general, persist.

We note that Figure~\ref{zoom_sommerfeld_standard_yukawa} shows the
\emph{minimal} increase due to the tower  (for a given value of
$\epsilon_\phi$). For smaller values of $\Lambda_{\uv}$, the increase in
the Sommerfeld enhancement can be larger and can even approach
$\mathcal{O}(1)$ values with $\epsilon_\phi\sim \Lambda_\ir/\Lambda_\uv$. This is easily seen by taking $\Lambda_{\uv} \sim \alpha_0M$ in
\eqref{inelastic_saturated_enhancement}. However, such small values of
$\Lambda_{\uv}$ require a more careful treatment with 
the sum evaluated numerically (to account for the full
$n$-dependence of the logarithms in \eqref{summ_approx}). This analysis bears out that the enhancement
increases for smaller values of $\Lambda_{\uv} \ll M_{Pl}$. Thus, if
the UV scale is $\Lambda_{\uv} \ll M_{Pl}$ a
greater enhancements result. Physically, the increase in the
enhancement results from the fact that the ratio of couplings in the
dual theory goes like 
\bea
\frac{\alpha_0^2}{\alpha_n^2}\sim \frac{\log^2(2 \Lambda_{\uv}/m_n)}{\log(\Lambda_{\uv} /\Lambda_{\ir})}\ \frac{1}{n}\quad \mathrm{for}\quad n>1\ .
\eea
This ratio increases with decreasing $\Lambda_{\uv} $. The heavier mediators therefore
couple more strongly to the scattering field as $\Lambda_{\uv} $ decreases, thereby
increasing their contribution to the Sommerfeld enhancement. Thus, we
can generalize the results by saying that, e.g.,  the
increase in the enhancement is $\gtrsim20\%$ for $\Lambda_{\uv} \lesssim
M_{Pl}$ when $\epsilon_\phi\sim 10^{-3}$. 

To summarize, for small values of $\epsilon_\phi\ll \epsilon_v$, the
Sommerfeld enhancement from a tower of mediators is larger than the
standard single-mediator saturated off-resonant result by a factor of $\sim
[\log(1/\epsilon_\phi)/\log(\Lambda_{\uv}/\Lambda_{\ir})]$:
\bea
S\rightarrow S_0\times \left[1+\frac{\log(1/\epsilon_\phi)}{\log(\Lambda_{\uv}/\Lambda_{\ir})}\right]\ ,
\eea
where $S_0=\pi/\epsilon_v$ is the small-$\epsilon_\phi$ off-resonant result due to a standard attractive Yukawa potential. This increase results from an effective increase in the coupling, $\alpha\rightarrow \sum_n\alpha_n$. For larger $\epsilon_\phi$ the exponential suppression in the Yukawa-potentials for the heavier modes is important and, in particular, the potential due to the heavier mediators becomes negligible relative to that of the lighter ones. The smooth transition from the regime where only the zero-mode is important to that where heavier mediators play a roll is seen clearly in Figure~\ref{zoom_sommerfeld_standard_yukawa}: for small $\epsilon_\phi\ll\epsilon_v$ the enhancement is larger due to the sum, but smoothly transitions to the standard value with increasing $\epsilon_\phi$.
\section{The Enhancement from a Tower: With a Mass Split\label{sec:split_sommer_tower}}
Thus far we have considered the Sommerfeld enhancement for a Dirac fermion (the elastic case).  Standard calculations show that  a small mass-splitting can increase the overall size of
the Sommerfeld effect, both on- and off-resonance, and shift the location of the resonances~\cite{Slatyer:2009vg}. In this section we comment on the Sommerfeld enhancement from a tower of mediators in the inelastic case, i.e.~the Dirac fermion is now split into two Majorana fermions with some small mass-splitting. Our presentation will be brief as our main points follow readily from the previous section. We limit our attention to a few key differences relative to the single-mediator analysis and follow the notations of Ref.~\cite{Slatyer:2009vg}.

Once mass splittings are introduced, the Sommerfeld enhancement  due to a single mediator is found by solving the coupled Schr\"odinger  equations~\cite{Slatyer:2009vg}
\bea
\chi''(x) = \mathcal{V}(x)\chi(x),
\eea
where $\chi(x)$ is now a two-component vector with the lightest (ground-state) fermion in the top position and
\bea
\mathcal{V}(x)=\left(
\begin{array}{cc}
-\epsilon_v^2&-e^{-\epsilon_\phi x}/x\\
-e^{-\epsilon_\phi x}/x& \epsilon_\delta^2-\epsilon_v^2
\end{array}
\right).
\eea
Here we consider an attractive Yukawa potential and restrict ourselves to $s$-wave annihilations for simplicity. We have also introduced the dimensionless variable $x=\alpha M r$ and the dimensionless quantities $\epsilon_v=v/\alpha$, $\epsilon_\phi=m/(\alpha M)$, and $\epsilon_\delta=\sqrt{2\delta/M}/\alpha$, where $\delta$ is the mass splitting between the two Majorana fermions. A detailed study of this system reveals that the enhancement for s-wave annihilation of ground state fermions can be cast as~\cite{Slatyer:2009vg}
\bea
S=\frac{2\pi}{\epsilon_v}\sinh\left(\frac{\epsilon_v\pi}{\mu}\right)\times
\left\{
\begin{array}{cc}
\left[\cosh(\epsilon_v\pi/\mu)-\cos(\sqrt{\epsilon_\delta^2-\epsilon_v^2}\pi/\mu +2\theta_-)\right]^{-1} & \epsilon_v<\epsilon_\delta\ ,\\
&\\
\frac{ \cosh\left[\left(\epsilon_v +\sqrt{\epsilon_v^2- \epsilon_\delta^2}\right)\pi/2\mu\right]\mathrm{sech}\left[\left(\epsilon_v -\sqrt{\epsilon_v^2- \epsilon_\delta^2}\right)\pi/2\mu\right]}{ \cosh\left[\left(\epsilon_v+ \sqrt{\epsilon_v^2- \epsilon_\delta^2}\right)\pi/\mu\right]-\cos2\theta_-}&\epsilon_v>\epsilon_\delta\ .
\end{array}
\right.
\eea
The factor $\theta_1$ contains an integral whose form is not important here, and $\mu\propto \epsilon_\phi/2$ (see Ref.~\cite{Slatyer:2009vg} for the precise expressions). We note the following limits, which are useful for our purposes. In the regime $\epsilon_\phi\ll\epsilon_v\ll\epsilon_\delta$, the enhancement becomes
\bea
S\rightarrow \frac{2\pi}{\epsilon_v},\label{mass_split_sommer}
\eea
which is larger than the elastic ($\delta=0$) result by a factor of two.  Also, in the limit $v\rightarrow0$ one finds that the enhancement saturates at~\cite{Slatyer:2009vg}
\bea
S=\left(\frac{2\pi^2}{\mu}\right)\times\left[1-\cos(\epsilon_\delta\pi/\mu-2\theta_-)\right]^{-1} >\frac{\pi^2}{\mu}.
\eea

We would like to understand the Sommerfeld enhancement due to a tower of mediators. 
Following the previous section, we define the dimensionless quantities in terms of the zero-mode mass and coupling (i.e. $m\rightarrow m_0$ and $\alpha\rightarrow \alpha_0$) and the Yukawa potentials in $\mathcal{V}(x)$ are replaced by a sum over mediators:
\bea
\frac{e^{-\epsilon_\phi x}}{x}\rightarrow \frac{1}{x}\left\{e^{-\epsilon_\phi x} +\sum_{n>0} \frac{\alpha_n}{\alpha_0}\ e^{-\delta_n\epsilon_\phi x}\right\}.
\eea
Based on the insights obtained in the elastic case, we can already deduce the following. For larger values of $\epsilon_\phi$ the enhancement should reduce to the standard expressions obtained with a single mediator. In this limit the exponential suppression of the heavier modes renders their contribution to the potential negligible relative to that of the zero-mode, and the latter essentially mimics a standard mediator. 

On the other hand, for small values of $\epsilon_\phi\ll\epsilon_v,\ \epsilon_\delta$, some of the exponentials are well approximated by their leading term, giving rise to an effective sum over Coulomb potentials, while others are exponentially suppressed. In this regime the standard analysis carries through, modulo the replacement $\alpha\rightarrow\bar{\alpha}=\sum_n\alpha_n$. For example, in the regime $\epsilon_v\gg \epsilon_\delta,\ \epsilon_\phi$ the enhancement is:
\bea
S\rightarrow \frac{2\pi}{\bar{\epsilon}_v}&=&\frac{2\pi}{\epsilon_v}\times \sum_n\frac{\alpha_n}{\alpha_0}\nonumber\\
&\simeq& \frac{2\pi}{\epsilon_v}\times \left\{1+ \frac{\log(1/\epsilon_\phi)}{\log(\Lambda_\uv/\Lambda_{\ir})}\right\}\nonumber\ .
\eea
As an example, with $\Lambda_{\ir}\sim$~GeV and $\epsilon_\phi\sim10^{-3}$ the value of the enhancement is\footnote{We give the numerical result.}
\bea
S&\rightarrow &\frac{2\pi}{\epsilon_v}\times 
\left\{1.2\right\}\quad\mathrm{for}\quad  \Lambda_\uv\sim M_{Pl} .
\eea
Once again the enhancement is increased by (at least) a factor of $\sim20\%$ relative to the standard result, with the precise increase dependent on the the UV scale $\Lambda_\uv$. Furthermore, compared to the standard off-resonant Sommerfeld enhancement with no mass splittings ($S\rightarrow \pi/\epsilon_v$ for $\epsilon_\phi\ll\epsilon_v$), the enhancement is increased by a factor $\gtrsim2.4$ for $\Lambda_\uv\lesssim M_{Pl}$ when both a mass-splitting and the tower of mediators is included.

In the limit $v\rightarrow0$, with small $\epsilon_\phi\ll\epsilon_\delta<1$, one obtains
\bea
S=\left(\frac{2\pi^2}{\bar{\mu}}\right)\times\left[1-\cos(\bar{\epsilon}_\delta\pi/\bar{\mu}-2\theta_-)\right]^{-1} >\frac{\pi^2}{\bar{\mu}},
\eea
where the ``barred'' quantities are obtained from their ``unbarred'' counterparts by the replacement $\alpha\rightarrow \bar{\alpha}$. This lower bound saturates at
\bea
S>\frac{\pi^2}{\bar{\mu}}=\frac{\pi^2}{\mu}\times \left\{1+ \frac{\log(1/\epsilon_\phi)}{\log(\Lambda_\uv/\Lambda_{\ir})}\right\}\ .
\eea
Thus we find that, relative to the single-mediator result, the saturated value of the Sommerfeld enhancement in the inelastic case is enhanced by the tower of mediators in both the small $\epsilon_\phi$ and the small $\epsilon_v$ limits. In the non-limiting cases the enhancement is readily found numerically by following the procedures outlined in Ref.~\cite{Slatyer:2009vg} with the modified potential due to the tower. 
\section{A Warped Model for a Tower of Mediators\label{sec:dual_model}}
We have studied the Sommerfeld enhancement due to a tower of mediators. This can arise in models where the scattering object is charged under a $U(1)$ factor that weakly-gauges a global symmetry of a strongly-coupled CFT (that confines in the IR). In this section we describe a  basic warped model that, via AdS/CFT, is dual to a 4D theory with a tower of vector mediators. This setup that was employed in the previous sections. 

Consider a
warped space with metric~\cite{Randall:1999ee}
\beq
ds^2= \frac{1}{(kz)^2}(\eta_{\mu\nu}dx^{\mu}dx^{\nu} -
dz^2)= G_{MN} dx^{M}dx^{N},
\label{bulkmetric}
\eeq
where $z \in [k^{-1},\,R]$ labels the extra dimension, $k$ is the curvature, and $\mu,\nu$
($M,N$) are the 4D (5D) Lorentz indices.  A 
$U(1)$ gauge symmetry propagates in the bulk of the 5D space and we include a UV-localized fermion $\psi$ with mass $M$ and  non-zero $U(1)$ charge. This field plays the role of the scattering/annihilating object that experiences a low-velocity Sommerfeld enhancement. 

In order to obtain an enhancement we require $MR\gg 1$, ensuring light KK vectors with masses $m_n\ll M$.  The IR-brane scale is identified with the confinement scale of the previous sections, $R^{-1}\Leftrightarrow \Lambda_{\ir}$, while the curvature $k$ plays the role of the UV scale $k\Leftrightarrow \Lambda_\uv$. Via the AdS/CFT correspondence~\cite{Maldacena:1997re}, RS models are considered dual to strongly-coupled 4D theories that are conformal over some range of energies but possess a mass gap in the IR~\cite{ArkaniHamed:2000ds}. The model that we consider, comprised of a UV-localized fermion that is charged under a bulk $U(1)$ symmetry, is dual to a 4D theory in which the CFT possesses a weakly-gauged global $U(1)$ symmetry, and the spectrum contains a tower of spin-one modes ($\rho_n$). The $U(1)$ gauge boson ($\gamma'$) kinetically-mixes with the CFT modes, like the mixing of $\rho$ and the photon in the SM~\cite{Agashe:2002jx}. The fermion is a fundamental field, external to the CFT, and  couples directly to $\gamma'$ due to its non-zero $U(1)$ charge. Although $\psi$ has no direct coupling to the CFT modes, such a coupling is induced by the $\gamma'-\rho_n$ mixing. Thus, in addition to processes like $\psi\bar{\psi}\rightarrow 2\gamma'$, annihilations into the CFT,  like $\psi\bar{\psi}\rightarrow \rho_n\rho_m$, can occur (kinematics permitting). As far as the Sommerfeld enhancement is concerned, scattering processes involving $\psi$ can be enhanced  at low-energies due to the exchange of virtual $\gamma'$ quanta, as per usual for a scattering object coupled to a mediator. However, in addition to $\gamma'$ exchange,  the tower of CFT modes can also be exchanged. As we have seen, regions of parameter space exist in which  the exchange of these CFT modes (dual to the exchange of KK vectors) does indeed increase the low-energy Sommerfeld enhancement. Note that this result depends on the fact that the dual theory 4D contains separate kinetic-mixing terms for $gamma'$ and each of the composites $\rho_n$.

The curvature $k$ can be considered as a free parameter whose value affects the coupling strength of $\psi$ to the KK vectors (see below). In the dual 4D theory the value of $k$ corresponds to the UV scale at which the strongly-coupled theory enters the conformal phase; equivalently, the conformal symmetry is broken by the UV cutoff at the scale $\sim k$. 
 Consistency therefore demands that we restrict $k$ to the range $M\lesssim k\lesssim M_{Pl}$. The lower bound ensures that the backreaction of $\psi$ can be neglected while the reason for the upper bound is obvious. For values of $k\ll M_{Pl}$, the inclusion of 4D Einstein gravity necessitates a large UV-localized Einstein-Hilbert term~\cite{George:2011sw}, while for $k\sim M_{Pl}$ the 4D Planck mass can be induced entirely by the bulk.\footnote{In the language of the dual 4D theory, for $k\ll M_{Pl}$ we include the 4D Planck scale as an input in the ``fundamental'' sector of the theory, while for $k\sim M_{Pl}$ CFT loops can generate the 4D Planck scale. The CFT loops also contribute to the 4D Planck scale in the former case, but are subdominant to the ``fundamental'' part of the Planck mass. The Planck mass is given by $M_{Pl}^2\simeq M_{\uv}^2+M_*^3/k$ with $M_{\uv}\sim M_{Pl}$ ($M_{\uv}\ll M_*\sim M_{Pl}$) in the former (latter) case~\cite{George:2011sw}.} In either case, 4D Einstein gravity is readily included in the low-energy theory. In our analysis we did not consider large hierarchies between the curvature and the 5D gravity scale (we took $k\sim M_*$), and focused mainly on values of $k\sim M_{Pl}$. 

The action for the UV-localized fermion $\psi$ with $U(1)$ charge\footnote{The specific value of the charge is not important as one can resale the bulk
gauge-coupling.} $Q=+1$ is
\bea
S\supset\int d^5x\sqrt{-g} \left\{\frac{i}{2}\bar{\psi}\Gamma^\mu D_\mu \psi - M\bar{\psi}\psi+ H.c.\right\}\delta(z-k^{-1}),
\eea
where $\Gamma^\mu$ are the curved-space Dirac matrices and $g_{\mu\nu}$ is the brane restriction of the metric. Integrating over the extra dimension gives
\bea
S\supset\int d^4x \left\{i\bar{\psi}\gamma^\mu \partial_\mu \psi
  +\sum_n e_{n}\bar{\psi}\gamma^\mu \psi \phi_\mu^{n}- M\bar{\psi}\psi\right\},\label{DM_action_tower}
\eea  
where $\gamma^\mu$ are 4D Dirac matrices and we have
expanded the covariant derivative. 
Here $\phi_n^{\mu}$ denotes the $n$th KK vector, which has mass $m_n$. We label the lightest vector as $n=0$ and refer to it as the zero-mode. It has mass $m_0\lesssim R^{-1}$, while the higher modes have masses $m_n\sim n\pi/R$ (see Appendix~\ref{app:vector_spectrum_one_throat} for a discussion of the KK spectrum). The effective 4D coupling between $\psi$ and the $n$th KK mode ($e_n$) is defined in terms of the
5D gauge coupling ($e_5$):
\bea
e_{0}&=&e_5\ f_0(k^{-1}) \simeq - \frac{e_5\ \sqrt{k}}{\sqrt{\log(2k/m_0)-\gamma}}\ ,\nonumber\\
e_{n}&=&e_5\ f_n(k^{-1}) \simeq -\frac{e_5\ \sqrt{k}}{[\log(2k/m_n)-\gamma]}\,(n+1/4)^{-1/2}\ ,~~~{n \geq 1}.
\label{DM_coupling_n_}
\eea
We can relate the effective 4D
coupling constants for modes with $n>0$ (and $m_n/k\ll1$) to that of the $n=0$ mode. This gives
\bea
e_{n}\simeq
\frac{1}{\sqrt{n}}\frac{e_{0}}{\sqrt{\log(2k/m_n)-\gamma}}\ ,\quad
n>0\ ,\label{brane_coupling_rel_single_}
\eea
revealing the coupling to the zero mode as dominant. The fine
structure constants are therefore related as $\alpha_n/\alpha_0\simeq [\log(2k/m_n)n]^{-1}$

The action \eqref{DM_action_tower}, together with the value of the masses $m_n\sim n\pi/R$ (see Appendix~\ref{app:vector_spectrum_one_throat}) and the coupling relations \eqref{DM_coupling_n_}, provide all the ingredients for the analysis of the previous sections. We made use of these results in  Section~\ref{sec:Somm_tower_no_spilt} to approximate the sum in Eq.~\eqref{small_ephi_limit} [see Eq.~\eqref{inelastic_saturated_enhancement}]. Specifically, Eq.~\eqref{brane_coupling_rel_single_} gives
\bea
\sum_n\frac{\alpha_n}{\alpha_0}\simeq \sum_n \frac{1}{n}\frac{\log(k R)}{[\log(2 k /m_n)]^2}.\label{summ_approx}
\eea
For $k\gg M$ one can neglect the $n$-dependence of the logarithms as we are only interested in modes with $m_n\lesssim M\ll k$, giving
\bea
\sum_n\frac{\alpha_n}{\alpha_0}\simeq \frac{1}{\log(k R)}\sum_n \frac{1}{n}.
\eea
Using this in Eq.~\eqref{small_ephi_limit} and approximating the sum as an integral gives Eq.~\eqref{inelastic_saturated_enhancement} (with $kR\rightarrow \Lambda_\uv/\Lambda_\ir$).

We note that the effective theory describing the UV-localized field $\psi$ is governed by the local 4D cutoff $M_*$, for which we have focused on $M_*\sim M_{Pl}$. One can trust the theory describing $\psi$ for energies $E\ll M_*$ and therefore calculations for the production/absorption of bulk-vector KK modes locally on the UV-brane are reliable. The KK modes themselves experience strong interactions in the bulk for scales $E\gg R^{-1}$, however, this does not alter the local UV physics. 
\section{Application to the Cosmic Lepton Excess\label{sec:cosmic_lepton}}
Given the interest in the Sommerfeld enhancement due to the cosmic-lepton-excess, it is appropriate to comment on the applicability of our results in this context, even if only to avoid perfidious implication. We restrict ourselves to a few brief comments as a more detailed analysis of DM in related scenarios will appear elsewhere.

It has been suggested that the cosmic lepton excess observed by PAMELA, \emph{Fermi}, and others, may be explained in terms of TeV-scale dark matter annihilating into a light mediator with mass $\lesssim $~GeV~\cite{ArkaniHamed:2008qn,Cirelli:2008id}. The boosted final-state mediators decay to the SM, and if the mediator mass is below the GeV scale, decays to anti-protons are forbidden, in accordance with the data. Decays to light SM fields, like electrons, muons or  pions, are allowed, subject to the details of the model. These light SM fields are highly boosted and thus provide a candidate explanation for the cosmic lepton excess. 

The present-day annihilation cross section required to explain the data is much higher [\mbox{$\gtrsim\mathcal{O}(10^2)$}] than the usual WIMP cross section, $\langle \sigma v\rangle\simeq 3\times 10^{-26}\mathrm{cm}^3/s$. The  Sommerfeld enhancement increases the present-day annihilation cross section, though it can be difficult to obtain a large-enough enhancement to ensure compatibility with the data (see, e.g.,~\cite{Meade:2009rb,Zavala:2009mi,Feng:2010zp} and references therein). If the DM has a mass-splitting, the situation is more promising due to the increased enhancement shown in \eqref{mass_split_sommer}~\cite{Finkbeiner:2010sm}. The existence of local substructure in the DM halo can also improve the fit to the data~\cite{Slatyer:2011kg}.  Note, however, that recent gamma-ray observations of the galactic center by HESS disfavor large regions of the parameter space, subject to the precise nature of the DM density profile\footnote{The HESS galactic-center background subtraction method precludes the attainment of limits on DM annihilations if the inner $\sim450~pc$ of the Milky Way contains a constant-density core~\cite{Abramowski:2011hc}.}~\cite{Abazajian:2011ak}. Also note that other astrophysical explanations for the lepton-excess exist (e.g.~\cite{Hooper:2008kg,Mertsch:2010fn}), some of which make predictions for the $B/C$ ratio expected in cosmic ray fluxes~\cite{Mertsch:2011gd}.

Given the increase in the Sommerfeld enhancement obtained with a tower of mediators, it is of interest to determine if this result could aid in accommodating the data. We will make some preliminary comments on the specific case considered here (a tower of mediators admitting a weakly-coupled dual description). The main point to make is that the application of these results would involve a non-standard DM scenario. Weakly-coupled warped models of the type considered here are dual to large-$N$ CFTs with $N$ estimated to be on the order of $N\simeq 4\pi(M_*/k)^{3/2}$~\cite{Gubser:1999vj}. The likely utility of this framework depends on whether the CFT modes are in equilibrium in the early universe. Clearly, if $k/M_*\ll1$ the number of degrees of freedom thermalized in the early universe can be dominated by the CFT, giving $g_*\sim N^2$. In this case the DM freeze-out cross section must be lower than the standard WIMP value in order to achieve the right abundance. In addition, the running of the $U(1)$ coupling constant can be modified due to CFT loops, which can reduce the low-energy coupling relative to models with less degrees of freedom. This exacerbates the need for a present-day Sommerfeld enhancement, which will have to exceed the ``standard'' value of $\sim\mathcal{O}(10^2)$. However, this feature is precisely that required if there are sizable amounts of local substructure. Indeed, the model falls under the category of ``new irrelevant channels'' according to the definitions of~\cite{Slatyer:2011kg}. On the other hand, if the DM is produced non-thermally by inflaton decays or some other such mechanism, as may be needed if the early universe is not reheated above the IR scale $\Lambda_{\ir}$, the CFT modes need not be in equilibrium in the early universe. In this case the typical DM annihilation cross section may remain on the order of the standard WIMP value, and the increased Sommerfeld enhancement could be useful in bringing the present-day annihilation cross-section in line with experimental requirements. 

There are a number of other details that could also come into play; for example, there would be extra annihilation channels involving higher KK modes, which would affect the thermal history in the early universe and the annihilation spectrum in the present-day. We will discuss these matters in a forthcoming work~\cite{bh_km}.
\section{Conclusion\label{sec:conc}}
We have studied the Sommerfeld enhancement due to a tower of
mediators, specializing to a system in which the scattering object is
charged under a weakly-gauged global symmetry of a
strongly-interacting (broken) CFT. This scenario can be motivated in
models of secluded dark matter in which the mediator scale is
generated by hidden-sector confinement. The resulting composites
contribute to the enhancement. We find that the exchange of multiple
mediators can increase the enhancement in the off-resonant region; for
a scattering object of  mass $M\sim$~TeV and a lightest mediator at
the GeV scale, the increase is found to be $\sim20\%$ for $\Lambda_\uv\sim
M_{Pl}$. The enhancement increases if the conformal symmetry is broken
in the UV
at scales $\Lambda_\uv\ll M_{Pl}$, due to the increased coupling strength of the
tower of mediators. A weakly-coupled dual model was employed in our analysis, though the general idea admits wider applicability.
\section*{Acknowledgements\label{sec:ackn}}
The author is most grateful to B.~v.~Harling and T.~Slatyer.
\appendix
\section{A Kaluza-Klein Tower of Mediators\label{app:vector_spectrum_one_throat}}
We briefly consider the KK spectrum of the bulk vector, following the hidden-vector discussion of~\cite{McDonald:2010iq}. The action for a bulk $U(1)$ symmetry on a slice of $AdS_5$ is  
\bea
S &\supset& 
- \frac{1}{4}\int d^4x\,dz\;\sqrt{G}\,G^{MA}G^{NB}\Phi_{MN}\Phi_{AB},
\eea
where $\Phi_{MN}$ is the field strength for the bulk vector $\Phi_M$. Symmetry breaking can be induced by an explicit
 Higgs  or by imposing a Dirichlet boundary 
condition on the IR brane~\cite{Csaki:2003dt}.  
We focus on the Higgsless case (in unitary gauge with $\Phi_5=0$).  The 5D gauge field is KK expanded as
\beq
\Phi_{\mu}(x,z) = \sum_nf_n(z)\phi^n_{\mu}(x),
\eeq
where the bulk wave functions satisfy~\cite{Davoudiasl:1999tf}
\beq
\left[z^2\partial_z^2 -z\partial_z + z^2m^2_n\right]f_n(z) &=& 0,\nonumber\\
\int  \frac{dz}{(kz)}\ f_n(z)\ f_m(z)&=&\delta_{nm}.
\eeq
The solutions, in terms of the Bessel functions $J_1$ and $Y_1$, are
\beq
f_n(z) = \frac{(kz)}{N_n}\left\{J_1(m_nz) +\beta_nY_1(m_nz)\right\}.
\label{wavefunc-gauge}
\eeq
A Neumann boundary condition at the UV brane ($z=k^{-1}$) leads to
\beq
\beta_n = -\frac{J_0(m_n/k)}{Y_0(m_n/k)} \simeq 
\frac{\pi}{2}\ \frac{1}{\log(2k/m_n)-\gamma},
\eeq
where the last expression holds for $m_n/k\ll1$ and $\gamma$ 
is the Euler-Mascheroni constant.
The eigenvalues $m_n$ are fixed by the Dirichlet
boundary condition in the IR, 
\beq
\beta_n=-\frac{J_1(m_n R)}{Y_1(m_n R)}~.
\label{dirichletfreq}
\eeq
 For $n$ greater than a few, the KK masses are well approximated by $m_nR\simeq (n+1/4)\pi$, while the mass of the lightest mode ($n=0$) is
\beq
m_0 \simeq \frac{1}{R}\sqrt{\frac{2}{\log(2kR)-\gamma}}.
\eeq
This mode has a wavefunction given by Eq.~\eqref{wavefunc-gauge} with
$N_0^{-1} \simeq \sqrt{2/kR^2}$. The UV-localized field $\psi$ couples to the KK modes with strength $e_{n}=e_5\ f_n(k^{-1})$, which depends on the UV values of the KK wave functions. These may be approximated as
\beq
f_0(k^{-1}) &\simeq &-\frac{\sqrt{k}}{\sqrt{\log(2k/m_0)-\gamma}}\ ,
\nonumber\\
f_n(k^{-1}) &\simeq &\frac{\sqrt{k}}{[\log(2k/m_n)-\gamma]}\ \frac{1}{\sqrt{n+1/4}}\ ,
~~~{n \geq 1}.
\label{epsilonn}
\eeq
For $m_n\ll k$, the coupling strength for the higher KK modes can be approximately related to that of the zero mode:
\bea
e_n\simeq \frac{1}{\sqrt{\log(2kR)}}\times \frac{e_0}{\sqrt{n+1/4}} \quad\mathrm{for}\  n>0.\label{App_coupling_rel}
\eea 
The ratio $e_n/e_0$ increases with decreasing $k$, and thus the higher modes couple more strongly as the UV scale is decreased. 

In addition, the particle spectrum
contains a tower of massive KK gravitons, whose spectrum is found 
by perturbing the background metric in the usual way~\cite{Davoudiasl:1999jd}. 
The massless zero-mode is the standard 4D graviton. Irrespective of the value of $k$, the couplings of the KK gravitons to the UV localized fermion $\psi$ are highly suppressed~\cite{George:2011sw} and play no direct role in our analysis. The gravitons couple to the KK vectors and thus influence their decay properties (see~\cite{McDonald:2010iq} for details) but are not important in this work.


\end{document}